\documentclass[journal]{IEEEtran}
\usepackage[T1]{fontenc}
\usepackage{cite}
\usepackage{url}
\usepackage{amsmath,amssymb,amsfonts}
\usepackage{graphicx}
\usepackage{textcomp}
\usepackage{xcolor}
\usepackage{hyperref}
\usepackage{bm}
\usepackage{amsthm}
\usepackage{enumerate}
\usepackage{booktabs}
\usepackage{diagbox}
\usepackage{supertabular}
\usepackage{subfigure}
\usepackage{caption}
\usepackage{algorithmicx}
\usepackage{comment}
\usepackage[linesnumbered,ruled,vlined]{algorithm2e}

\begin{document} 

\title{
%Generative AI for Net-Zero Carbon Mobile Networks: Insights, Architecture, Outlooks
Generative AI for Low-Carbon Artificial Intelligence of Things with Large Language Models
}

% Carbon-friendly打算换成net-zero carbon

\author{Jinbo Wen, Ruichen Zhang, Dusit Niyato, \textit{Fellow, IEEE}, Jiawen Kang, Hongyang Du,\\ Yang Zhang, and Zhu Han, \textit{Fellow, IEEE}
\thanks{
%This work was supported by the National Natural Science Foundation of China (NSFC) under Grants No. 62102099, No. U22A2054, and No. 62071343, Guangzhou Basic Research Program under Grant 2023A04J1699, Guangdong Basic and Applied Basic Research Foundation under Grant 2023A151514 0137, Foundation of State Key Laboratory of Public Big Data under Grant No. PBD2023-12, and Collaborative Innovation Center of Novel Software Technology and Industrialization. This work was also supported by the National Research Foundation, Singapore, and Infocomm Media Development Authority under its Future Communications Research \& Development Programme, Defence Science Organisation (DSO) National Laboratories under the AI Singapore Programme (AISG Award No: AISG2-RP-2020-019 and FCP-ASTAR-TG-2022-003), and the Singapore Ministry of Education (MOE) Tier 1 (RG87/22).
J. Wen and Y. Zhang are with the College of Computer Science and Technology, Nanjing University of Aeronautics and Astronautics, China (e-mails: jinbo1608@163.com; yangzhang@nuaa.edu.cn). 

R. Zhang, D. Niyato, and H. Du are with the School of Computer Science and Engineering, Nanyang Technological University, Singapore (e-mails: ruichen.zhang@ntu.edu.sg; dniyato@ntu.edu.sg; hongyang001@e.ntu.edu.sg). 

J. Kang is with the School of Automation, Guangdong University of Technology, China (e-mail: kavinkang@gdut.edu.cn). 

Z. Han is with the Department of Electrical and Computer Engineering, University of Houston, USA (e-mail: hanzhu22@gmail.com).

%\textit{*Corresponding authors: Yang Zhang and Jiawen Kang.}
}
}

\maketitle

\begin{abstract}
By integrating Artificial Intelligence (AI) with the Internet of Things (IoT), Artificial Intelligence of Things (AIoT) has revolutionized many fields. However, AIoT is facing the challenges of energy consumption and carbon emissions due to the continuous advancement of mobile technology. Fortunately, Generative AI (GAI) holds immense potential to reduce carbon emissions of AIoT due to its excellent reasoning and generation capabilities. In this article, we explore the potential of GAI for carbon emissions reduction and propose a novel GAI-enabled solution for low-carbon AIoT. Specifically, we first study the main impacts that cause carbon emissions in AIoT, and then introduce GAI techniques and their relations to carbon emissions. We then explore the application prospects of GAI in low-carbon AIoT, focusing on how GAI can reduce carbon emissions of network components. Subsequently, we propose a Large Language Model (LLM)-enabled carbon emission optimization framework, in which we design pluggable LLM and Retrieval Augmented Generation (RAG) modules to generate more accurate and reliable optimization problems. Furthermore, we utilize Generative Diffusion Models (GDMs) to identify optimal strategies for carbon emission reduction. Numerical results demonstrate the effectiveness of the proposed framework. Finally, we insightfully provide open research directions for low-carbon AIoT.
\end{abstract}


% GAI本身exist carbon emission % How to estimate carbon emission reduced by GAI % 中间穿插一下IAI和RAG

\section{Introduction}
%As the amount of data generated by businesses and consumers worldwide increases, the energy consumed by cloud data centers can lead to significant carbon emissions due to the bandwidth used to transfer data to the cloud. 
Currently, Artificial Intelligence of Things (AIoT) is ushering in a new era of the digital economy by supporting the technological revolution in many fields\cite{liu2023enabling}, such as smart healthcare and smart agriculture\cite{wen2023generative}. However, the impact of AIoT on energy consumption and carbon emissions is a topic of concern\cite{li2023carbon}. Specifically, the advent of transformative technologies such as AI-Generated Content (AIGC), the Internet of Things (IoT), and Metaverse has led to a significant surge in data volume within AIoT\cite{lai2023resource}. According to research conducted by Transforma Insights, the global deployment of edge devices is projected to rise from 2.7 billion to 7.8 billion in the next decade. Furthermore, the broader category of IoT-connected devices is expected to surpass 30 billion worldwide by 2025, and the mobile data traffic of a mobile device will reach 257.1 GB per month by 2030, which is a substantial increase of 50 times compared to the data volume in 2010\cite{lai2023resource}. Hence, the rapid growth in power consumption of edge loads and the scarcity of energy resources pose significant energy challenges to AIoT, resulting in new environmental impacts. It is worth noting that low energy consumption and low carbon emissions are related but distinct concepts in terms of environmental impacts. Specifically, 
\begin{enumerate}[1)]
\item {\textit{Low energy consumption:}} Its goal is to decrease the total energy usage of a system by using energy-efficient technology. The energy usage includes both renewable and non-renewable sources, which helps to conserve natural resources and reduce dependence on fossil fuels. 
%This refers to the use of energy-efficient technology to complete tasks or maintain a certain level of activity. It focuses on reducing overall energy demand, including renewable energy, which helps to conserve natural resources and reduce dependence on fossil fuels.
\item {\textit{Low carbon emissions:}} It focuses on reducing the carbon footprint, particularly from burning fossil fuels such as coal and oil. Low carbon emissions include choosing renewable energy sources such as wind, solar, and hydro, which have minimal or no direct carbon emissions, even if they are not the most energy-efficient.
\end{enumerate}

Generative AI (GAI) is a branch of AI technology that can produce various types of content, including text, imagery, and audio\cite{wen2023generative, du2023beyond}. The demand for AIGC services spanning various domains is driven by the advancement in GAI models. For instance, DALL$\cdot$E 2\footnote{\url{https://openai.com/dall-e-2}}, developed by OpenAI, possesses the capability to generate original and realistic images based on user prompts consisting of textual descriptions. ChatGPT\footnote{\url{https://chat.openai.com/}}, as a transformer-based Large Language Model (LLM), has showcased its remarkable capability in textual content generation tasks\cite{zhang2024interactive}. In addition to data interpretation, GAI can generate synthetic data critical to users and networks\cite{wen2023generative}, which enables predictive actions according to network condition changes using past and synthetic data, ensuring efficient network resource allocation and minimizing energy consumption from network operations. Thanks to these prominent capabilities, GAI has been explored to reduce energy consumption in many domains, such as manufacturing, transportation, and agriculture.
\begin{comment}
\begin{itemize}
    \item \textbf{Manufacturing.} GAI is achieving sustainability in manufacturing by continuously monitoring energy consumption patterns and adjusting machine operations accordingly. For example, GAI can throttle the machine when energy demand is low, thus reducing energy consumption.
    \item \textbf{Transportation.} GAI is transforming the transportation industry by optimizing operations. For example, GAI can improve the energy efficiency of charging schedules for electric vehicles, thus reducing the energy consumption of transportation operations.
    \item \textbf{Agriculture.} Agriculture is undergoing a revolution empowered by GAI. For example, by harnessing image recognition and data analysis, GAI can detect crop disease, achieve personalized crop management, and enhance resource efficiency, thus ushering in innovative and sustainable agriculture.
\end{itemize}
\end{comment}

%Thanks to these prominent capabilities of GAI, including resource allocation optimization\cite{lai2023resource}, optimal network decision\cite{wen2023generative}, and personalized user interaction\cite{du2023age}.
\begin{figure*}[t]
%\vspace{-0.5cm}
\centerline{\includegraphics[width=0.95\textwidth]{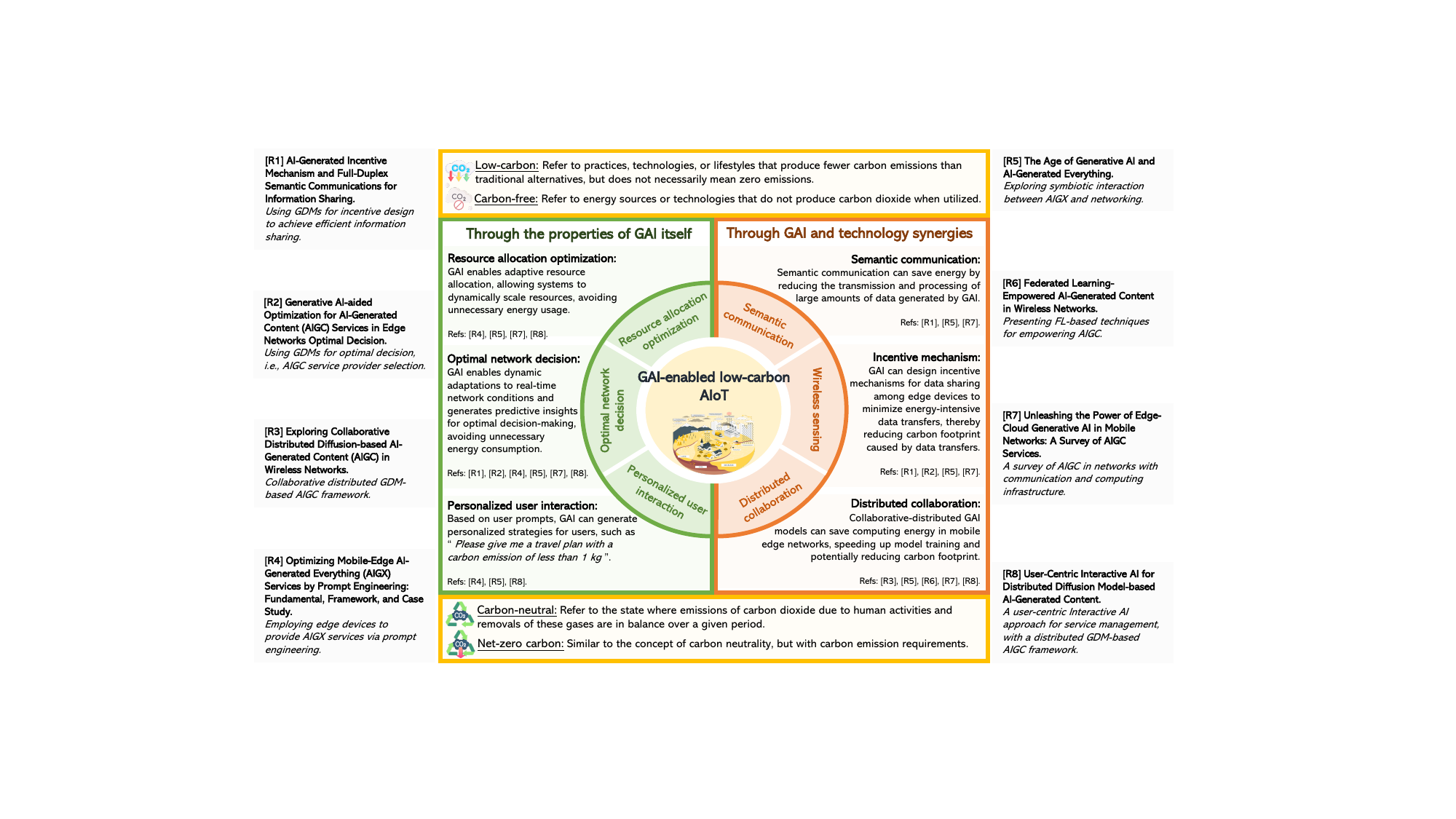}}
\captionsetup{font=footnotesize}
\caption{A brief summary of recent studies on GAI and intelligent networking. We introduce the concepts of common carbon emission goals and focus on exploring the potential of GAI to enable low-carbon AIoT from two perspectives, either through the properties of GAI itself or through the synergy of GAI with other techniques.}
\label{related}
\end{figure*}

Current research commonly focuses on using Discriminative AI (DAI) to reduce network carbon emissions\cite{li2023carbon, hussin2023systematic}. However, DAI which focuses primarily on analyzing or classifying existing data has a poor capability of adapting to the dynamic environment of AIoT. Inspired by the revolutionary capabilities of GAI, GAI-driven solutions hold the immense potential to optimize energy consumption and reduce carbon emissions in AIoT. In addition, GAI can cope with the dynamic changes of network conditions and adaptively adjust optimal strategies without retraining, avoiding additional carbon footprints. Therefore, GAI opens up new avenues to achieve low-carbon AIoT. Unlike traditional green mobile networks, which primarily focus on reducing energy consumption and enhancing energy efficiency, \textit{low-carbon AIoT enabled by GAI focuses on utilizing GAI to minimize carbon emissions and promotes sustainable practices across the entire network ecosystem}, which has the following potential characteristics:
\begin{itemize}
    \item \textbf{Renewable energy integration:} Low-carbon AIoT prioritizes the integration of renewable energy sources, utilizing advanced techniques such as GAI-driven energy harvesting and optimization algorithms to efficiently harvest renewable energy\cite{rane2023contribution}, such as solar and wind energy, thereby minimizing the dependence on energy production based on fossil fuel.
    \item \textbf{Intelligent network management:} Low-carbon AIoT utilizes intelligent network management techniques such as GAI-driven network management\cite{wen2023generative} to effectively monitor network energy consumption in real-time, and dynamically optimize resource distribution to minimize carbon emissions.
    \item \textbf{Green network infrastructure:} Low-carbon AIoT emphasizes the application of environmental infrastructure components, such as sustainable materials and energy-efficient hardware, and utilizes advanced techniques, e.g. GAI-driven optimal Intelligent Reflection Surface (IRS) deployment\cite{naeem2023joint}, to effectively optimize network infrastructure and reduce carbon emissions.
\end{itemize}

Figure \ref{related} presents recent advances in integrating GAI with intelligent networks and related concepts of carbon emission goals, including low-carbon, carbon-free, carbon-neutral, and net-zero carbon. \textit{To the best of our knowledge, this is the first work that systemically provides forward-looking research on the potential of GAI enabling low-carbon networks, which is the first step toward carbon-neutral and net-zero carbon paradigms}. Our main contributions are summarized as follows:
\begin{itemize}
    \item We first investigate the main carbon emission impacts of mobile networks, then briefly discuss the limitations of DAI in carbon emission reduction, and systematically introduce GAI techniques, including their features and abilities to reduce carbon emissions.
    \item We explore the potential applications in GAI enabling low-carbon AIoT by penetrating the mobile network architecture, i.e., Energy Internet (EI), data center networks, and mobile edge networks.
    \item We propose an LLM framework combining Retrieval Augmented Generation (RAG) for carbon emission optimization, where we design pluggable LLM and RAG modules that rely on knowledge bases and context memory to generate carbon emission optimization problems.
    \item We adopt Generative Diffusion Models (GDMs) to identify optimal strategies for carbon emissions. Simulation results of a real carbon emission optimization case study demonstrate the effectiveness of the proposed framework.
    %and exploring the potential applications of GAI-enabled carbon-friendly mobile networks from the physical layer, network layer, and application layer.
\end{itemize}

\begin{table*}
  \centering
  \caption{The Illustration of Carbon Emissions in Mobile Technology.}
  \begin{tabular}{c}
    \includegraphics[width=0.9\textwidth]{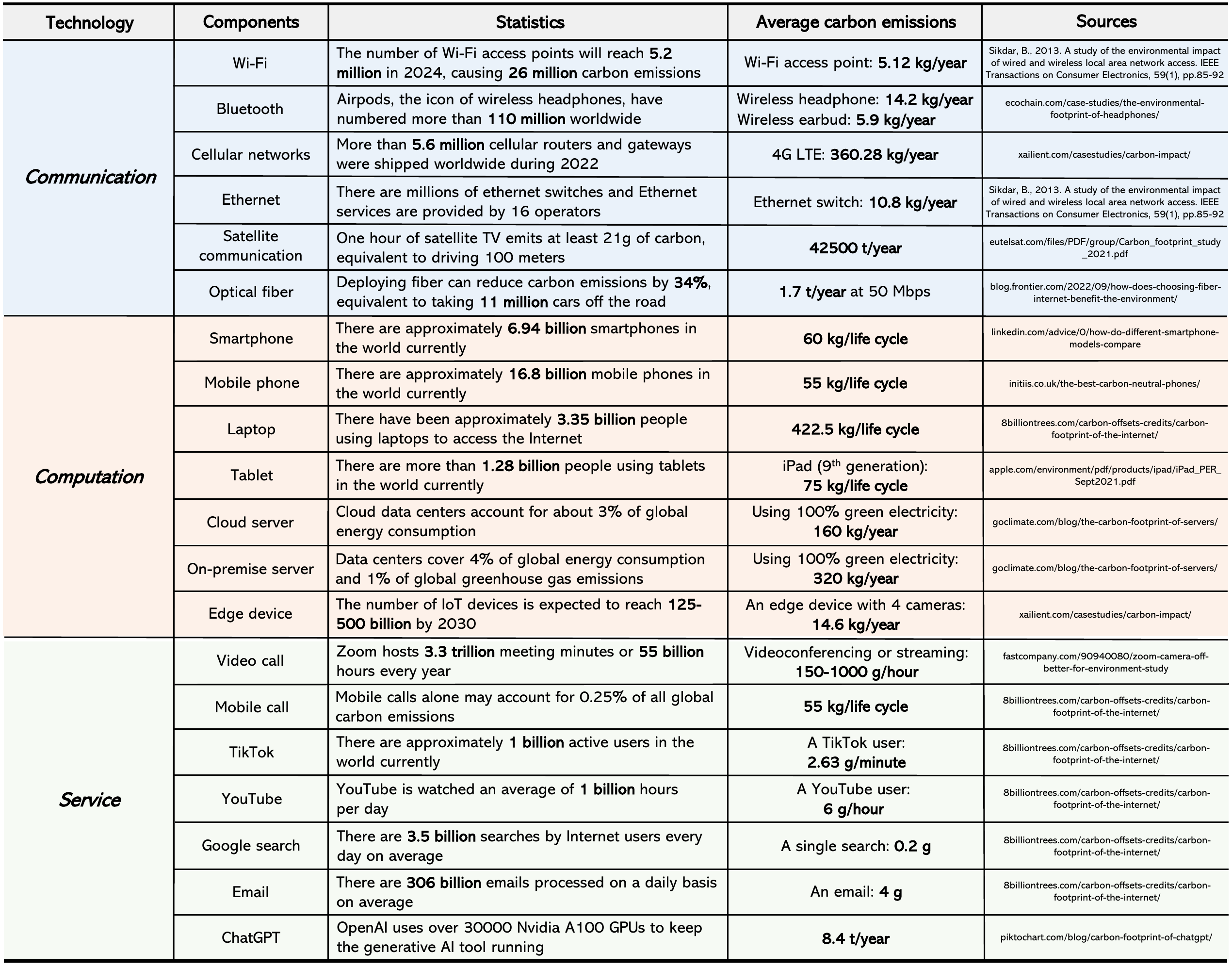} \\
  \end{tabular}
  \label{Energy}
\end{table*}

%reduce energy consumption associated with carbon footprint in mobile edge networks

\section{Motivations for Low-Carbon AIoT Using Generative AI}
In this section, we first discuss the carbon emission impact of mobile networks. Then, we systematically introduce GAI techniques, involving their basic architectures and potential applications in reducing energy consumption associated with carbon footprint. Finally, we briefly review recent studies on GAI and networking, exploring the potential ability of GAI to enable low-carbon AIoT.

\subsection{Carbon Emission from Mobile Networks}
Multi-access Edge Computing (MEC) technologies have emerged to bring computational resources closer to mobile devices. The shift from cloud to edge computing solves the limitations of high service latency and bandwidth consumption by processing data at the edge network\cite{wen2023freshness}. However, due to the increased energy usage and distributed infrastructure of mobile networks\cite{liu2023enabling}, moving the computation process to the edge will further exacerbate the carbon emission of AIoT. As shown in Table \ref{Energy}, we investigate the carbon emissions of mobile technology, including communication, computation, and service technologies. Since AIoT and mobile networks are symbiotic and mutually beneficial\cite{liu2023enabling}, we discuss the main carbon challenges of mobile networks from the perspectives of communication and computation in the following part.
%leading to lower energy consumption and reduced carbon emissions compared to cloud computing, 
%mobile edge computing brings critical energy consumption challenges.

%\subsubsection{Energy consumption challenges in mobile edge computing}
%\emph{\textbf{C1. Sensing energy consumption:}} Sensing devices in mobile edge networks are responsible for collecting data from the environment. The collected sensing data serving as input facilitates complex operations such as data analysis and processing, allowing for data-driven decision-making. However, the influences of sensing mechanisms, sampling rates, and sampling quality lead to energy consumption during the sensing process\cite{mao2023green}. For example, compared to an optical sensor, an imaging sensor can consume higher power used for heart rate monitoring\cite{mao2023green}. Moreover, with the rapid development of Internet of Things, the large-scale adoption of mobile edge computing will lead to a surge in sensing data, potentially increasing overall energy consumption and additional carbon emissions in mobile edge networks.

\emph{\textbf{C1. Communication impact on carbon emissions:}} 
%\subsubsection{Trade-off of offloading performance and carbon emission reduction}
%By leveraging the capabilities of mobile edge computing, offloading computationally intensive tasks such as machine learning to the network edge can achieve low-carbon computation\cite{ma2023towards}, significantly reducing energy consumption and carbon emissions associated with data transmission and processing. However, due to the inherent limitations of edge resources, the inference accuracy achieved in the edge server is usually worse than that achieved in the cloud server\cite{ma2023towards}. Therefore, striking an optimal balance between minimizing inference accuracy loss and realizing carbon emission reduction emerges as a critical concern.
When mobile devices communicate with edge servers, they need to transmit collected data over wireless channels for data processing and analysis tasks, leading to communication energy consumption. 
%The communication energy consumption relies on various factors, including carrier frequency, modulation schemes, and antenna deployment\cite{mao2023green}. For instance, in a 2.4GHz WiFi network, an end-user device can consume up to 32 mW\cite{mao2023green}.
The current mobile network has larger bandwidths and more antennas, dramatically increasing energy consumption and carbon emissions\cite{li2023carbon}. According to rough estimates\footnote{\url{https://www.rcrwireless.com/20220923/5G}}, China's 5G network generates more than 60 million tons of carbon emissions nationwide every year. Besides, satellite communication requires significant energy consumption for satellite operations and ground infrastructure, leading to a notable environmental impact on mobile networks. For example, the satellite fleet causes 37,484 tons of carbon emissions every year. Therefore, it is crucial to develop appropriate communication technologies based on GAI that meet the needs of applications while minimizing carbon emissions.

\emph{\textbf{C2. Computational impact on carbon emissions:}}
While benefiting human productivity and efficiency, the large-scale use of computational devices has led to the explosion of data and computation in mobile networks, resulting in huge carbon emissions. For example, there were 7.7 billion mobile phones in use worldwide in 2020, producing about 580 million tons of carbon emissions, equivalent to about 1\% of total global emissions. In mobile networks, the limited power capacity of edge devices may also present challenges in affording substantial computation energy consumption required for computational-intensive tasks, especially for AI model training and inference\cite{lai2023resource}. For instance, the energy consumption for training a ResNet-110 model\footnote{\url{https://builtin.com/artificial-intelligence/resnet-architecture}} on the NVIDIA Jetson TX2 platform amounts to approximately 8 × 105 Joules of energy. 

%Due to the inherent limitations of edge resources, the inference accuracy achieved in the edge server is usually worse than that achieved in the cloud server\cite{ma2023towards}. 
%Therefore, striking an optimal balance between minimizing inference accuracy loss and realizing carbon emission reduction also emerges as a critical concern.

In summary, it is necessary to enable AIoT to be low-carbon while ensuring system performance, thereby achieving sustainable development in intelligent fields such as smart cities and generative IoT \cite{wen2023generative}.

\subsection{Discriminative AI in Carbon Emission Reduction}
As a class of AI that aims to distinguish between different classes in a given dataset, DAI has been utilized in many specific tasks for reducing carbon emissions, such as renewable energy harvest\cite{li2023carbon}, carbon capture\cite{hussin2023systematic}, and energy management\cite{li2023carbon}. For example, the authors in \cite{li2023carbon} proposed a machine learning model to effectively coordinate the working state of 5G cells and avoid carbon efficiency traps. However, due to the dynamic and heterogeneous nature of AIoT\cite{wen2023generative}, DAI has obvious limitations in terms of carbon reduction:
\begin{itemize}
    \item \textbf{Limited applicability:} The efficacy of DAI in carbon emission reduction heavily depends on specific applications\cite{hussin2023systematic}, such as renewable energy integration, transportation optimization, and smart grid management. When new applications emerge, the existing DAI models need to be retrained, resulting in huge energy consumption and carbon emissions.
    \item \textbf{Training data availability:} DAI models require significant data to train and optimize to find effective patterns and trends for energy conservation\cite{li2023carbon}. However, when the training data is of low quality or unavailable, it may be challenging for DAI models to correctly identify inefficiencies and provide effective suggestions. Consequently, the availability and quality of the training data can be a potential limitation.
    \item \textbf{Resource requirement:} Implementing DAI models at scale for energy efficiency can be costly\cite{li2023carbon}, requiring computational resources, data storage and processing infrastructure, and even technical expertise. Besides, the continuous maintenance and upgrading of the model may add additional costs and carbon emissions.
\end{itemize}

\begin{figure*}[t]
%\vspace{-0.5cm}
\centerline{\includegraphics[width=0.95\textwidth]{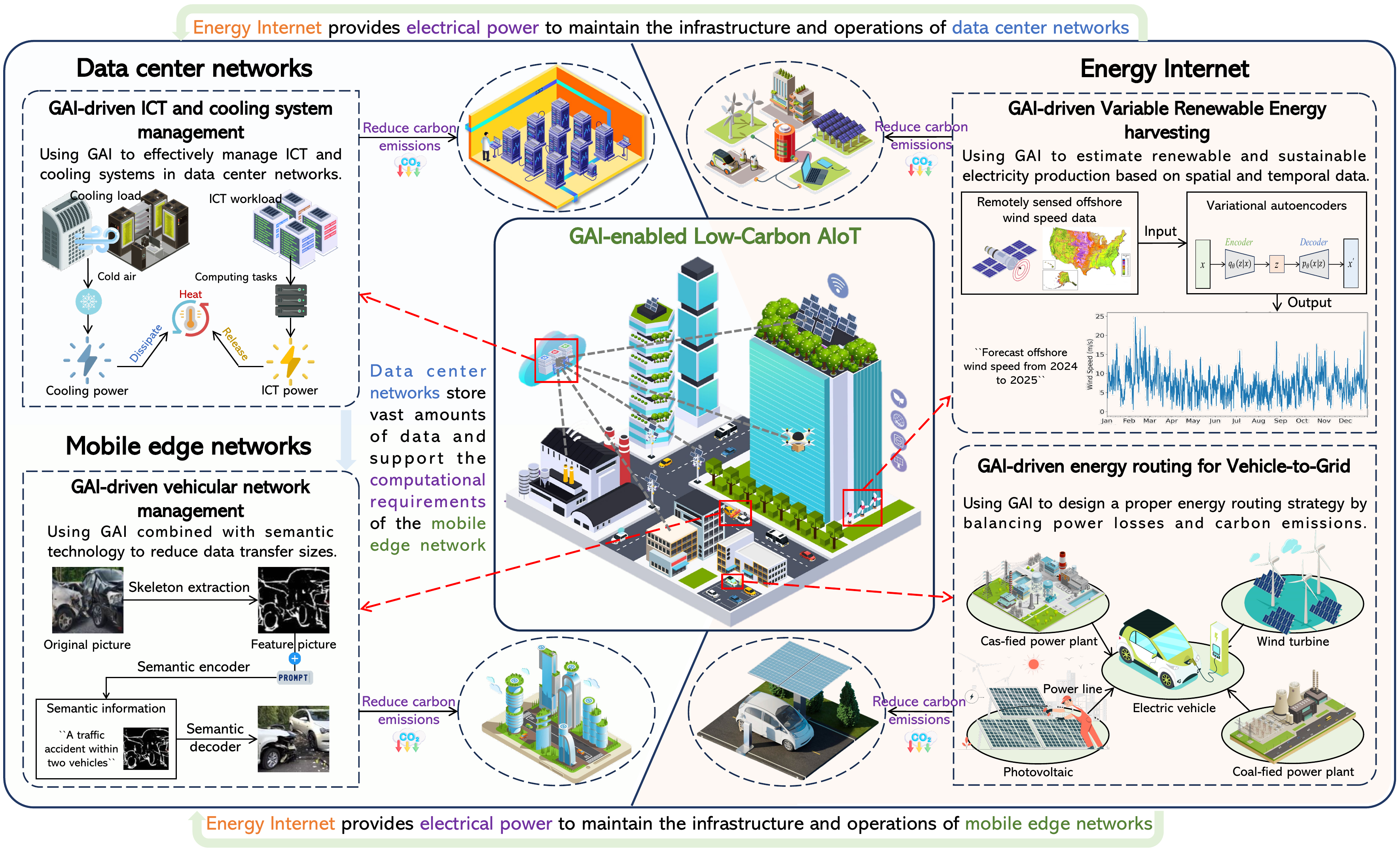}}
\captionsetup{font=footnotesize}
\caption{Potential mobile network applications in low-carbon AIoT. We study how GAI can reduce the carbon emissions of EI, data center networks, and mobile edge networks to enable low-carbon AIoT. In particular, EI can provide electrical power to maintain the infrastructure and operations of data center networks and mobile edge networks, and data center networks can provide computational resources to mobile edge networks.}
\label{system}
\end{figure*}

\subsection{Generative AI Technologies and their Relations to Carbon Emissions}
Unlike the focus of DAI on detecting existing patterns, GAI focuses on generating new data samples, holding significant capabilities of content creation, data augmentation, and even network resource optimization\cite{wen2023generative}. The foundations of GAI technology and their relations to carbon emissions are discussed as follows:
\begin{itemize}

    \item \textit{Generative Adversarial Networks (GANs):} GANs consist of generator and discriminator networks, where the generator network aims to generate new data and the discriminator network aims to distinguish synthetic data from real data\cite{wen2023generative}. Since the two networks engage in iterative training and competition, GANs possess data generation and discrimination capabilities. For carbon emission reduction, GANs, such as BiLSTM-CNN-GAN\footnote{\url{https://typeset.io/questions/what-are-the-bilstm-cnn-gan-algorithm-3wt75zl0t7}}, can predict energy consumption and carbon emissions, facilitating efficient resource management and planning.
    %The two networks are trained together, interacting and competing against each other, thereby continuously improving both performances\cite{cao2023comprehensive}.
    %With good performance in generating realistic samples\cite{liu2023deep}, GANs can be utilized not only for data augmentation but also for IoT anomaly detection\cite{cook2019anomaly}. Notably, unlike traditional AI methods that typically require retraining on labeled data to adapt to changes, GANs can learn the underlying data distribution in an unsupervised manner, enabling the adaptation to evolving anomalies without explicit labeling.
    
    %GANs can learn the normal patterns and behaviors of IoT devices or sensor data for anomaly detection in IoT applications
    
    %facilitating them more computationally efficient in mobile edge networks\cite{xu2023unleashing}.
    \item \textit{Retrieval Augmented Generation:} RAG is an advanced technique for enhancing the reliability and accuracy of GAI models by retrieving facts from an external knowledge base\cite{zhang2024interactive}, which can augment user prompts by adding relevant retrieved data in context to allow LLMs to generate accurate answers.
    %\textit{Transformers} are core techniques of LLMs, learning complex dependencies and generating coherent contextual instances based on the self-attention mechanism. 
    Especially, LLMs supported by RAG can generate accurate carbon emission optimization strategies by accessing external databases, such as documents about carbon emission reduction.

    \item \textit{Generative Diffusion Models:} 
    GDMs 
    %which are mainly formulated into three categories, i.e., denoising diffusion probabilistic models, score-based generative models, and noise conditional score networks. 
    consist of forward diffusion and denoising processes inspired by non-equilibrium thermodynamics theory\cite{wen2023generative}, gradually transforming initial random samples into the target output through several iterative denoising steps.
    %In the forward diffusion process, original images are gradually destroyed by the successive addition of Gaussion noise. Then, GDMs are trained to denoise the images to learn the inverse diffusion process for the construction of desired images. 
    With the incredible capability of image generation, GDMs have the potential to be applied to optimize image generation tasks, ensuring a more sustainable use of computing resources and reducing carbon emissions.
    
    %Unlike GANs, GDMs do not require adversarial training, which avoids training instabilities and ensures model scalability.

    \item \textit{Other GAI techniques:} \textit{Variational Autoencoders (VAEs)} can represent data in a probabilistic latent space, which enhances accuracy in short-term energy forecasting and optimizes energy usage for carbon emission reduction. \textit{Flow-based Generative Models (FGMs)} facilitate data generation by transforming input data distributions from simple to complex through a series of reversible transformations, which can be potentially applied to predict weather patterns, ensuring that as much renewable energy as possible is harvested to reduce carbon emissions.
    %Unlike VAEs and GANs, FGMs can explicitly learn the data distribution, and their loss functions are the negative log-likelihood, which use the back-propagation algorithm for gradient computation during data generation\cite{xu2023unleashing}, 
    %Unlike VAEs and GANs, FGMs possess the distinctive capability to learn explicitly the data distribution and directly compute the probability density function during generation \cite{du2023beyond}. Therefore, FGMs can circumvent resource-intensive computation and directly model complex probability distributions, which can be effectively applied in IoT domains such as traffic flow optimization\cite{xu2023unleashing} and anomaly detection in network traffic.
\end{itemize}

\begin{table*}
  \centering
  \caption{Comparison between Traditional and GAI Methods in Mobile Network Applications for Low-Carbon AIoT.}
  \begin{tabular}{c}
    \includegraphics[width=0.9\textwidth]{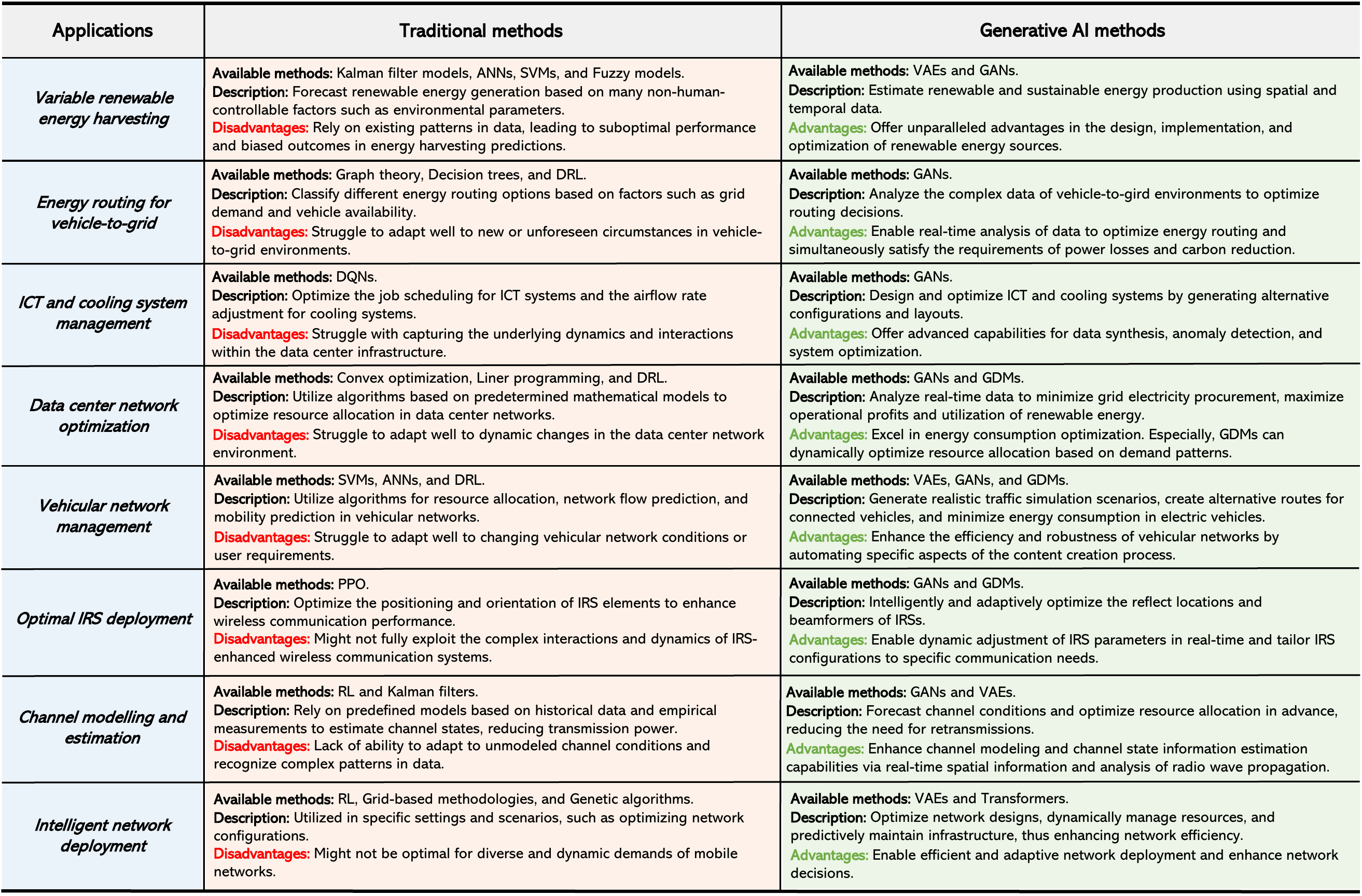} \\
  \end{tabular}
  \label{Comparison}
\end{table*}

\section{Generative AI for Low-Carbon AIoT}
%In this section, we study how GAI can enable mobile networks to be carbon-friendly by permeating and influencing the components of mobile networks, namely the power system, the data center, and the mobile edge network.
In this section, we study how GAI can reduce the carbon emissions of mobile network components, namely EI, data center networks, and mobile edge networks, thereby enabling low-carbon AIoT, as shown in Fig. \ref{system}.

\subsection{Energy Internet}
%\subsubsection{GAI-driven semantic communications}
%By converting messages into semantic information for transmission\cite{du2023age}, Semantic Communication (SemCom) promises to significantly alleviate the scarcity of communication resources in mobile edge networks. However, existing SemCom systems lack the ability for context-aware reasoning, which hinders their potential in enhancing communication reliability and efficiency\cite{xia2023generative}. Fortunately, GAI, with the incredible abilities of context-reasoning and cross-modal content synthesis\cite{wen2023generative}, can enhance semantic context-reasoning and achieve higher spectrum utilization\cite{xia2023generative}, leading to reduced energy consumption and carbon emissions. For instance, the authors in \textbf{\emph{[R1]}} of Fig. \ref{related} proposed an efficient information-sharing scheme for full-duplex device-to-device semantic communication, enabling users to avoid repetitive and heavy computational tasks. To motivate users to share semantic information, the authors formulated an incentive mechanism based on contract theory and utilized GAI, i.e., the diffusion model, to design optimal contracts. For a given environment state, comparative analyses demonstrate that the proposed AI-generated contract algorithm outperforms the DRL-based schemes.

EI can provide reliable and efficient power supplies to maintain the operations of mobile edge networks and data center networks. In response to the continuous increase of anthropogenic carbon emissions in EI, GAI is considered a powerful technology to achieve low-carbon EI, which in turn optimizes mobile network performance.
\subsubsection{GAI-driven Variable Renewable Energy (VRE) harvesting} 
%Renewable-powered mobile networks have attracted significant attention from both academia and industry\cite{ma2023towards}. 
In EI, the integration of renewable energy within residential areas can enhance energy supplies for community-level mobile edge networks, which reduces carbon emissions and mitigates air pollution. In \cite{rane2023contribution}, the authors highlighted the significant role of GAI in advancing renewable and sustainable energy technologies. Specifically, by analyzing real-time data on environmental conditions and network demands, GAI can intelligently adjust energy harvesting mechanisms\cite{rane2023contribution}. For instance, GAI can optimize the positions of solar panels installed for base stations to capture the maximum amount of solar energy during sunny days with ample sunlight. In addition, GAI can efficiently estimate renewable and sustainable electricity production using spatial and temporal data from other renewable energy sources, such as biomass and onshore wind energy. This estimation can be used to schedule network and computing workloads to reduce reliance on carbon-based power generation. Thus, GAI-driven VRE harvesting holds immense potential to mitigate environmental impacts and enhance energy accessibility in mobile networks.

\subsubsection{GAI-driven energy routing for Vehicle-to-Grid (V2G)}
As a typical EI scenario, V2G technology integrates Electric Vehicles (EVs) into the power grid. However, the main power source of EVs does not rely only on the power grid, but also on a variety of other energy sources, such as renewable energy sources\cite{hua2023carbon}. In this case, the energy is transmitted throughout the networked EI scenario, like information routing on the Internet\cite{hua2023carbon}. It is worth noting that power losses and carbon reduction are mutually exclusive in energy transmission\cite{hua2023carbon}. As a result, how to design a proper energy routing strategy to simultaneously satisfy these two targets is significant. GAI has been applied in designing routing strategies\cite{dong2021generative}. Combined with DRL, GAI can analyze complex data sets to optimize routing decisions for reducing unnecessary carbon emissions.

\subsection{Data Center Networks}
Data center networks play a pivotal role in storing, processing, and managing vast amounts of data, which supports the computational requirements of mobile edge networks. However, data center networks are carbon-intensive due to their massive energy consumption\cite{cao2022toward}. To reduce the carbon emission of data center networks, we explore the adoption of GAI for Information and Communication Technology (ICT) and cooling system management and network optimization.

\subsubsection{GAI-driven ICT and cooling system management}
The electricity from ICT and cooling systems accounts for about 86\% of the total energy consumption of data center networks\cite{cao2022toward}. In the data center network, the ICT system generates heat during operation, and the cooling system is designed to dissipate this heat to maintain a suitable for the equipment. Thus, the effective management of both ICT and cooling systems is essential to enhance energy efficiency and reduce carbon emissions in data center networks\cite{cao2022toward}. GAI is expected to optimize the management of ICT and cooling systems in data center networks. For instance, by analyzing real-time data from ICT and cooling systems, GAI can predict their equipment failures in advance\cite{wen2023generative}, optimizing the corresponding maintenance activities and reducing energy wastage.

%With extensive data flow across multiple devices in mobile networks\cite{lai2023resource}, especially in the Internet of Vehicles (IoV) where vehicles continuously exchange data, data center network management is crucial to ensure the efficient network operation and reduce carbon emissions in mobile networks\cite{cao2022toward}. Given the incredible capabilities of data representation and generation prowess\cite{wen2023generative}, GAI can optimize network management by adaptively allocating resources based on real-time data, proactively predicting network congestion, and even integrating with semantic technology to reduce data transfer sizes. For instance, the authors in \cite{du2023beyond} illustrated the application of GDMs combined with semantic technology in IoV design. Besides, the authors addressed vehicle-to-vehicle resource allocation and formulated a Quality of Experience (QoE) metric grounded in received image fidelity and the transmission rate. Numerical results demonstrate that the proposed GDM-based scheme outperforms DRL schemes under the same parameter settings.

\subsubsection{GAI-driven network optimization}
%GAI-driven network optimization, a key element in achieving carbon-friendly mobile edge networks, leverages advanced GAI to 
%Optimizing network operations holds the potential to achieve energy efficiency improvement and carbon emission reduction. 
The network optimization for carbon-free data center networks primarily focuses on minimizing grid electricity procurement, maximizing operational profits, and maximizing utilization of renewable energy\cite{cao2022toward}. By analyzing real-time data encompassing user demands, network conditions, and energy availability, GAI excels in efficient resource allocation and energy consumption optimization\cite{lai2023resource}. In particular, GDMs can dynamically optimize resource allocation based on demand patterns, ensuring efficient utilization and minimizing energy consumption. For instance, 
%the authors in \cite{lai2023resource} first proposed the concept of \textit{generative mobile edge networks} by integrating GAI with mobile edge networks. Then, the authors performed a comprehensive case study focusing on resource-constrained mobile edge networks. Specifically, 
the authors in \cite{lai2023resource} proposed a Stackelberg game for efficient resource allocation and the objective of this game is to balance energy consumption and network performance. Then, they applied GDMs to find the optimal solution. 
%Numerical results reveal that the proposed GDM-based approach outperforms the conventional DRL approach, recording a $19.2\%$ increase in the edge server utility.

\begin{figure*}[t]
%\vspace{-0.5cm}
\centerline{\includegraphics[width=0.95\textwidth]{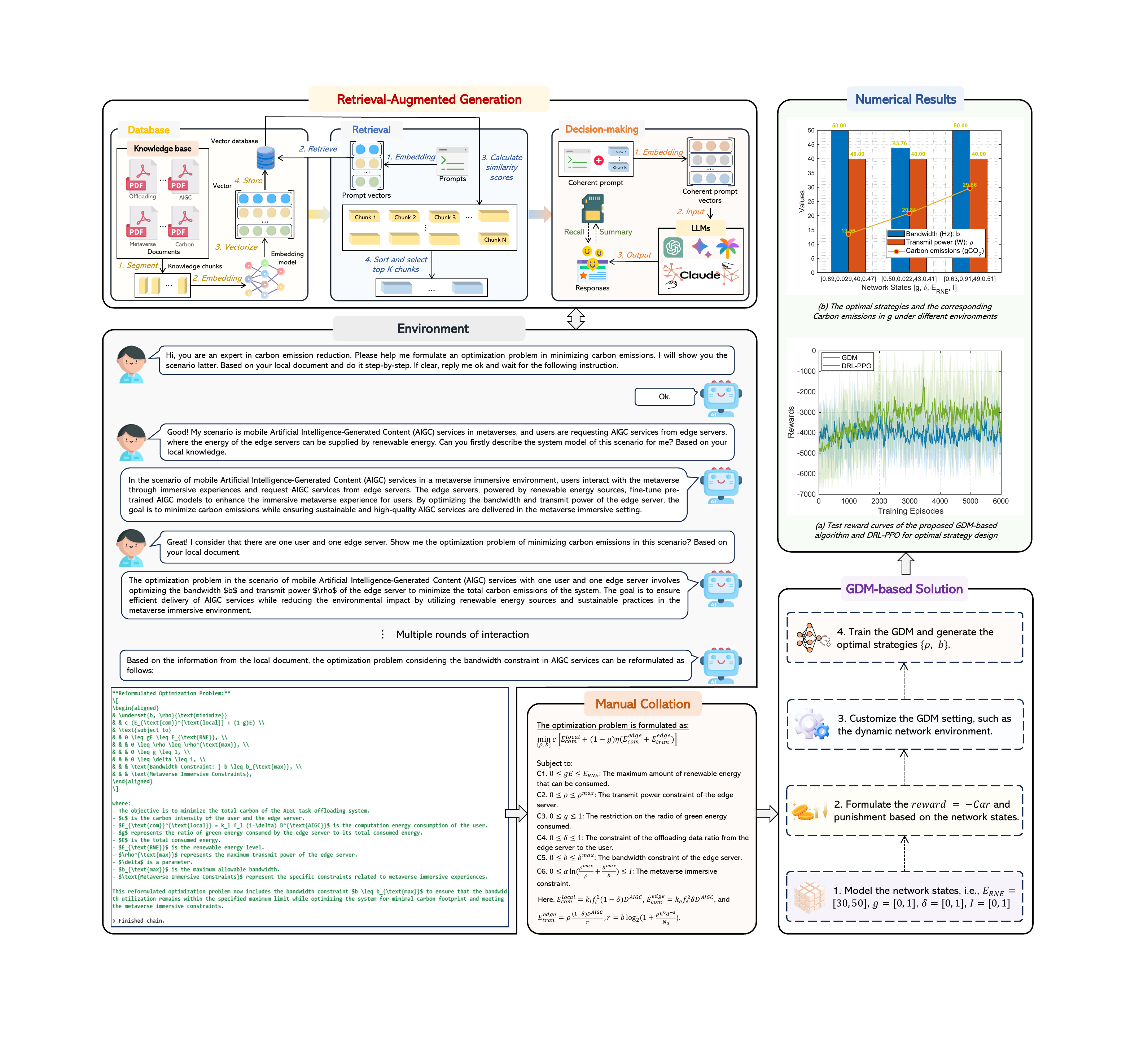}}
\captionsetup{font=footnotesize}
\caption{The LLM-enabled carbon emission optimization framework supported by RAG. In the proposed framework, RAG consists of three key components, i.e., database, retrieval, and decision-making, enabling accurate and reliable carbon emission optimization problems. Then, GDMs can be employed to generate optimal strategies, where the motivations and the specific process of utilizing GDMs for determining optimal strategies are introduced in \cite{du2023beyond}. }
\label{framework}
\end{figure*}

\subsection{Mobile Edge Networks}
%\subsubsection{GAI-driven energy-efficient resource allocation} 
With the large-scale adoption of MEC capabilities, mobile edge networks face environmental issues that conflict with global sustainable development goals\cite{li2023carbon}. Next, we study how GAI can permeate and influence the physical architecture of mobile edge networks to mitigate their environmental impacts.

%\subsubsection{GAI-driven antenna trajectory optimization} 
%By harnessing the capability of optimal decision, GAI can optimize the trajectory of antennas to maximize Spectral Efficiency (SE) or Energy Efficiency (EE)\cite{du2023age}. 
%The optimization of antenna positions at the transceiver and reflector can dynamically enhance channel conditions, maximizing Spectral Efficiency (SE) or Energy Efficiency (EE) and lowering energy consumption\cite{zheng2023flexible}. In contrast to DAI which uses historical data to determine the optimal trajectory based on specific objectives, GAI can perform dynamic trajectory planning based on environmental factors such as user locations and channel conditions, ensuring optimal coverage for users while minimizing energy consumption, thus efficiently utilizing limited resources and reducing the carbon footprint of mobile edge networks. For instance, the authors in \cite{zheng2023flexible} utilized the GDM to generate efficient antenna trajectories for reducing power consumption and enhancing SE and EE. Compared to conventional DRL algorithms, GDMs can achieve higher total EE. To strike a balance between SE and EE, the designed trajectory ensures that all antennas converge towards the center of the user population, thereby maximizing the overall EE.
\subsubsection{GAI-driven vehicular network management}
With extensive data flow across vehicles in the Internet of Vehicles (IoV), 
%especially in the Internet of Vehicles (IoV) where vehicles continuously exchange data, 
vehicular network management is crucial to ensure efficient communication and reduce carbon emissions in mobile edge networks\cite{cao2022toward}. Given the capabilities of data representation and generation prowess\cite{wen2023generative}, GAI can optimize network management by adaptively allocating resources based on real-time data, proactively predicting network congestion, and even integrating with semantic technology to enhance the efficiency and robustness of vehicular networks. For instance, the authors in \cite{du2023beyond} illustrated the application of GDMs combined with semantic technology in IoV design. Besides, the authors addressed vehicle-to-vehicle resource allocation, thereby reducing energy consumption while ensuring image fidelity and transmission performance.

%and formulated a Quality of Experience (QoE) metric grounded in received image fidelity and the transmission rate. 
%Numerical results demonstrate that the proposed GDM-based scheme outperforms DRL schemes under the same parameter settings.

\subsubsection{GAI-driven optimal IRS deployment}
IRSs have the capability of significantly improving energy efficiency and spectrum utilization with low-power and low-cost hardware\cite{naeem2023joint}. By leveraging GAI, the deployment of IRSs becomes intelligent and adaptive. The placement and configuration optimization of IRSs can ensure that IRS panels efficiently reflect wireless signals toward desired areas or users, which reduces the need for excessive signal transmission power, thus improving network performance and minimizing carbon emissions. For instance, the authors in \cite{naeem2023joint} focused on the joint optimization of the placement and reflecting beamforming matrix in the IRS-assisted 6G network. Specifically, the authors proposed a GAN-based DRL framework to jointly optimize the reflect locations and beamformers of IRSs. 
%Finally, the numerical results demonstrate that the proposed GAN-based DRL scheme can efficiently learn the optimal IRS deployment strategy in complex and dynamic wireless communication networks.

For clarity, the comparison between the traditional and GAI approaches applied in mobile network applications for low-carbon AIoT is summarized in Table \ref{Comparison}.

\section{LLM-enabled Carbon Emission Optimization Framework Supported by RAG}
In this section, we propose an LLM-enabled carbon emission optimization framework supported by RAG. We conduct a case study on carbon emission optimization for mobile AIGC task offloading in a metaverse environment and utilize GDMs to generate optimal strategies.

\subsection{Motivation}
%As shown in Fig. \ref{case}, \textit{part A} shows a Mobile Edge Computing (MEC) system with $N$ users and a base station deployed with MEC servers, where the MEC server has sufficient computing resources to perform many tasks offloading in parallel\cite{ma2023towards}. Within a period, users request AIGC services from the base station. Then, MEC servers fine-tune GAI models and execute inferences based on the user-provided prompts\cite{wen2023freshness}, and deliver AIGC services to users, e.g., generated images or videos. However, the process results in significant carbon emissions. Therefore, we focus on minimizing the total carbon emission of AIGC services while satisfying the user requirements of service latency. Note that the energy required for the base station is provided by both the power grid and renewable energy sources\cite{ma2023towards}.
Carbon emission optimization is a significant approach for minimizing environmental impacts from AIoT. It specifically refers to optimizing various sectors of mobile systems, such as data transfer energy efficiency\cite{li2023carbon}, data centers\cite{cao2022toward}, and task offloading\cite{ma2023towards}. Inspired by the exceptional decision-making capability of LLMs, we propose an LLM-enabled carbon emission optimization framework supported by RAG. By interpreting the network environment, the proposed framework can automatically formulate significant carbon emission optimization problems through simple interactions with network designers. With the support of RAG\cite{zhang2024interactive}, the framework can significantly lower the risk of human errors, improve the accuracy of problem formulation\cite{zhang2024interactiveS}, and speed up the design process by fusing the carbon emission reduction knowledge learned from the comprehensive knowledge base. Compared with the existing energy management strategies\cite{ma2023towards}, the generated strategy from the problem formulated by the RAG-supported LLM agent is more comprehensive and practical.

\subsection{Framework Design}
%The total system energy consumption is mainly composed of the computation energy consumption (denoted as $EE$) for GAI fine-tuning and inference and the communication energy consumption (denoted as $\sum_{n=1}^N ET_{b,n}$) for delivering AIGC services to users \cite{ma2023towards}. For the communication energy consumption, when the transmit power of the base station is large, although the AIGC service delay is greatly reduced, it will increase the communication energy consumption and cause a large amount of carbon emissions. On the contrary, the quality of services will be affected. Therefore, we intend to optimize the transmit power of the base station to minimize the total carbon emissions of the system (denoted as $C$) while guaranteeing the AIGC service delay. By using the Interactive AI (IAI) framework proposed in \cite{zhang2024interactive}, the appropriate optimization problem aligning with the ground truth can be formulated, as shown in \textit{Part B} of Fig. \ref{case}. Then, we utilize the Proximal Policy Optimization (PPO) method to demonstrate the effectiveness of the proposed scheme.
As shown in Fig. \ref{framework}, the environment under consideration represents the real-world scenarios described by the network designer. The interaction starts with an initial request from the network designer for assistance, and the LLM agent can generate decision-making results from RAG. The augmentation process of RAG is presented as follows:

\textbf{Database.} The RAG database is a large-scale knowledge base that involves a wealth of searchable academic texts\cite{zhang2024interactive}, such as academic papers on carbon reduction from IEEE Xplore. RAG takes the knowledge from the knowledge base and segments it into knowledge chunks. Then, these chunks are transformed into dense vector representations by embedding models and stored in a vector database for embedding search.

\textbf{Retrieval.} The requests of the network designer are first transformed into dense vector representations that are readily interpretable by the LLM agent\cite{zhang2024interactive}. Then, RAG retrieves relevant information from the vector database and calculates the similarity scores of these knowledge chunks. Finally, RAG sorts and selects the previous most similar chunks as the component of extended context prompts. 

\textbf{Decision-making.} Based on the request of the network designer and the selected chunks, LLM, such as ChatGPT, Gemini, or Bard, can formulate responses due to their reasoning and decision-making capabilities. Furthermore, these responses are stored in a repository, enabling the LLM agent to effectively recall and apply previous strategies when dealing with similar tasks\cite{zhang2024interactive}.

Upon the optimization problem generated by the LLM, the network designer can simply collate the generated optimization problem according to the requirements. Specifically, the network designer can customize subjective constraints and determine experimental parameters based on real scenarios. This process is almost burdenless. Since GDMs show superior performance in handling high-dimensional and complex optimization problems\cite{du2023beyond}, the network designer can leverage GDMs to generate optimal strategies and implement these strategies in real-world scenarios, thereby effectively achieving carbon emission reduction in AIoT. The technical principle and specific process of GDMs for solving optimization problems can be found in \cite{du2023beyond}. Note that the trained GDM can adapt to different states of AIoT systems to generate the optimal strategy for carbon emission reduction. 

In summary, the proposed LLM-enabled carbon emission optimization framework can help the network designer consider more factors for carbon emission reduction. In addition, RAG assists LLM agents to perform accurate inference and reduce the carbon emissions caused by inference tasks of LLMs, and prompt engineering can be applied within RAG to further enhance the interaction between the network designer and the LLM agent, allowing for precise information retrieval and generation based on finely tuned prompts.

\subsection{Case Study: Mobile AIGC Task Offloading in a Metaverse Environment}
To explore the effectiveness of the proposed framework, we conduct a case study on mobile AIGC task offloading in a metaverse environment, where mobile AIGC refers to the integration of AIGC with mobile edge networks\cite{wen2023freshness}.

\subsubsection{Scenario description} In the scenario of mobile AIGC services in a metaverse environment, users request AIGC services from edge servers, such as personalized avatars that provide users with immersive experiences in the metaverse. Edge servers, powered by renewable energy sources\cite{ma2023towards}, fine-tune pre-trained AIGC models and execute inferences to enhance the quality of immersive experiences for the users. To reduce service latency, AIGC tasks can be collaboratively executed by the edge servers and users, where the intermediate results of AIGC tasks are sent to users by the edge servers. In particular, we consider a user and an edge server in this scenario. By optimizing the bandwidth and transmit power of the edge server, the goal is to minimize the carbon emissions of an AIGC task through the offloading mechanism while ensuring high-quality AIGC services in the metaverse setting.

\subsubsection{Framework configuration} In our experiments, we call the GPT-4 model through the OpenAI API to implement the pluggable LLM module, and the RAG module is built on top of LangChain\footnote{\url{https://www.langchain.com/}}. We set the chunk size, chunk overlap, and retrieval results are set as $1000$, $200$, and $4$, respectively. Thus, the LLM agent can generate accurate models with a minimum number of retrieved tokens, i.e., a total of $4000$.

\subsubsection{Numerical results}We perform experiments by using PyTorch on NVIDIA GeForce RTX 3080 Laptop GPU. The numerical result module (a) of Fig. \ref{framework} shows test reward curves of the proposed GDM-based algorithm and Proximal Policy Optimization (PPO) for optimal strategy design. We can observe that the GDM achieves higher test rewards than the PPO, indicating better performance. The reason is that GDMs generate optimal strategies by diffusion process that can mitigate the impacts of noise and randomness\cite{du2023beyond}. To demonstrate that training GDMs does not result in excessive carbon emissions, we use a Python package called CodeCarbon\footnote{\url{https://github.com/mlco2/codecarbon}} and estimate the power consumption and carbon emissions caused by GDM training to be $8.148$ Wh and $1.672$ g, respectively. The numerical result module (b) of Fig. \ref{framework} illustrates optimal strategies and the corresponding carbon emissions under different network environments. We can observe that due to the exploration experience during the denoising, GDMs can determine the optimal strategy for low carbon emissions.

% Then, MEC servers fine-tune GAI models and execute inferences based on the user-provided prompts\cite{wen2023freshness}, and deliver AIGC services to users, e.g., generated images or videos. However, the process results in significant carbon emissions. Therefore, we focus on minimizing the total carbon emission of AIGC services while satisfying the user requirements of service latency. Note that the energy required for the base station is provided by both the power grid and renewable energy sources\cite{ma2023towards}.

\section{Future Directions}
\subsection{Carbon Emission Minimization Problems for Cloud-Edge-Device Architectures}
For cloud-edge-device architectures, one of the current problems of carbon emission minimization is the complexity of optimizing energy usage while considering dynamic workloads. To address this problem, future research can utilize GAI to dynamically adjust resource allocation and workload distribution based on changing environments.

%The transportation sector stands out as a significant contributor to carbon emissions. Although the transition to electric vehicles holds promise in mitigating carbon emissions, the effective reduction of carbon emissions within vehicular networks remains a complex challenge. Therefore, future research can explore the integration of GAI with deep learning models to accurately predict carbon emissions from traffic vehicles, enhancing the support development of proactive emission reduction strategies.

%\subsection{Integration of Generative AI and Mobile Edge Networks}
%The cloud-based AIGC service may introduce considerable service latency, which is unsuitable for specific mobile AIGC tasks requiring real-time immersion and exceptionally low service latency. Therefore, future research can explore the integration of GAI and mobile edge networks, emerging a novel generative edge network that possesses powerful capabilities of content generation and decision-making.

\subsection{Generative AI-enabled Carbon Trading through the Agent}
Carbon training is the buying and selling of credits that permit an entity to emit a certain amount of carbon dioxide. However, the opacity of carbon trading may cause additional carbon emissions. Therefore, future research can utilize GAI to facilitate the development of smart contracts for carbon trading, thereby ensuring transparency and security of carbon trading records on the blockchain.

%\subsection{Low-Carbon Space-Air-Ground-Integrated Networks (SAGINs)}
%The SAGIN is a comprehensive network architecture that integrates ground, air, and space communications to achieve ubiquitous connectivity. With the increasing concerns about environmental impacts, the need to enable SAGINs to be low-carbon is emerging. Therefore, future research can utilize GAI to enable low-carbon SAGINs by optimizing resource allocation and network architectures within SAGINs.

\subsection{Training Optimization for Generative AI Models}
Training models are the most energy-intensive phase of GAI. For example, training a large language model, such as OpenAI's GPT-4 or Google's PaLM, is estimated to lead to 300 tons of carbon emissions. Therefore, future research can explore techniques to optimize the training process for GAI models at the edge, such as federated learning, transfer learning, and distributed training algorithms, thereby reducing energy consumption and carbon footprint while maintaining GAI model performance.

\subsection{Carbon-aware Deployment of Generative AI Models} 
The existing centralized AIGC framework experiences significant service latency issues, resulting in the limited scalability of GAI applications\cite{wen2023freshness}. Therefore, future research can investigate the potential of edge computing and distributed architectures for GAI. Specifically, we can explore GAI model deployment at the edge in a carbon-aware manner, which can minimize the need for extensive data transfer and centralized cloud computing, leading to lower energy consumption and reduced carbon emissions.

\section{Conclusion}
In this article, we presented the prospect of GAI for low-carbon AIoT. First, we investigated the carbon challenges of mobile networks and systematically reviewed GAI techniques and their relationship with carbon emission reduction. Then, we explored the potential applications of GAI in reducing the carbon emissions of mobile network components for enabling low-carbon AIoT. Inspired by the outstanding capabilities of LLMs, we proposed an LLM-enabled carbon emission optimization framework supported by RAG, thus generating more accurate and reliable carbon emission optimization problems. Furthermore, we utilized GDMs to generate optimal strategies for carbon emission reduction. To validate the effectiveness of the proposed framework, we conducted a case study on mobile AIGC task offloading in a metaverse environment. Numerical results demonstrate that our LLM agent can generate precise carbon emission optimization problems with the minimum number of retrieved tokens, and the performance of GDMs for optimizing carbon emissions is $17.97\%$ higher than that of DRL-PPO. Finally, we discussed potential research directions that can further achieve low-carbon AIoT.

\bibliographystyle{IEEEtran}
\bibliography{ref}

\end{document}